\documentstyle[aps,eqsecnum,preprint]{revtex}
\tightenlines
\begin{document}

\title{Fluctuation-induced forces in critical fluids}
\author{M. Krech \\ Institut f\"ur Theoretische Physik, RWTH Aachen,
52056 Aachen, Germany}
\maketitle

\begin{abstract}
The current knowledge about fluctuation induced long-ranged forces is
summarized. Reference is made in particular to fluids near critical
points, for which some new insight has been obtained recently. Where
appropiate, results of analytic theory are compared with computer
simulations and experiments.
\noindent
\end{abstract}
\draft

\section{Introduction}
Forces between particles are governed by fields which themselves can be
considered as composed of particles mediating the interaction by continuous
exchange processes. Most prominent in the macroscopic world are
electromagnetic fields and gravitational fields. For simplicity we specialize
to electromagnetic forces here, but the line of argument scetched out in the
following is valid in general. Macroscopic bodies exert electromagnetic
forces on one another whenever they are charged, but if they are neutral all
electromagnetic forces apparently vanish. H.B.G. Casimir \cite{Casimir} 
was the first to realize that this is not quite correct, because the
electromagnetic field is fluctuating. These fluctuations may be due to quantum
fluctuations at zero temperature in vacuum or due to thermal fluctuations
in a cavity which is in contact with a heat bath. In any case the fluctuation
spectrum, i.e., the energies which are associated with the Eigenmodes of
the system and the form of the Eigenmodes themselves are manifestly
influenced by the {\em geometry} of the system. The geometry is given by
an arrangement of surfaces which impose boundary conditions on the fluctuating
field and thus determine its mode spectrum. The free energy, which contains
all information about the thermodynamic properties of the system, is
essentially given by a sum over all modes and therefore the free energy will
become geometry dependent. If, for example, two uncharged metallic bodies are
placed at a certain distance in vacuum the free energy of the
configuration depends on the shape of the bodies and the distance between
them. Therefore, there will be an {\em effective force} acting between the
bodies, which is given by the derivative of the free energy with respect
to their distance. Note that this force is a direct consequence of the
influence of the bodies on the electromagnetic fluctuation spectrum. Apart
from the macroscopic length scales set by the geometry there are no other
length scales in the system which limit the maximum wavelength of the
fluctuations and therefore the force is governed by powers of the imposed
length scales and scaling functions of their ratios, i.e., the resulting
force is {\em long - ranged}. Due to the history of their discovery
\cite{Casimir} these forces are now known as Casimir forces and the influence
of boundaries on the functional form of the free energy is known as the
{\em Casimir effect}. It should also be noted that the presence of additional
bodies in the above setup modifies the force between any two of them, i.e.,
it is not possible to express the Casimir effect as a sum of two - body
contributions only.

\subsection{A classic example}
There is a vast body of literature on the various aspects of the Casimir
effect in electromagnetism which are beyond the scope of this article. For
a summary we refer to review articles dedicated to these subjects
\cite{Reviews} and for a recent survey in a more general context we
refer to Chapter 3 of Ref.\cite{MKbook}. However, for the understanding of
the mechanism and the interpretaion of the results it is instructive to
demonstrate some of the fundamental physical concepts by a simple example,
which we call {\em classic} here for its historic meaning.

Let us assume that we have two parallel perfectly conducting plates placed
at a distance $L$ in vacuum. We further assume that the plates have an
infinite lateral extension so that we consider the thermodynamic limit
with respect to the surface area $\cal A$ of the plates and only discuss
the free energy $\cal F$ per unit area, i.e., we calculate $F =
\lim_{{\cal A} \to \infty} {\cal F} / {\cal A}$. Note, that $L$ is the
only macroscopic length scale in the problem. The mode spectrum of the
electromagnetic field in this parallel plate geometry is particularly
simple. In three dimensions the wave vector ${\bf q} = ({\bf p}, k_n)$
consists of a lateral component ${\bf p} = (p_x, p_y)$ which is unconstrained
by the geometry and a discrete perpendicular component $k_n = n\pi / L$ for
$n = 1, 2, 3, \dots$ due to the condition that the electric field vector at
each of the metallic surfaces must be aligned with the surface normal. A
single mode is then characterized by ${\bf p}$ and $n$ and its energy level
spacing is given by
\begin{equation}
\label{epspn}
\varepsilon_{{\bf p},n} = \hbar c \sqrt{{\bf p}^2 + (n\pi / L)^2}.
\end{equation}
The energy content $E_{{\bf p},n}$ of a particular mode is given by its
occupation number $m_{{\bf p},n} = 0, 1, 2, \dots$ in the form
$E_{{\bf p},n} = \varepsilon_{{\bf p},n} \left( m_{{\bf p},n} +
{1 \over 2} \right)$. The free energy per unit area is then given by (see
also Ref.\cite{KKSM98} for a recent reconsideration of the Casimir effect at
general temperature)
\begin{equation}
\label{F}
F = \int^\Lambda {d^2p \over (2\pi)^2} \sum_{n=1}^N \varepsilon_{{\bf p},n}
+ 2 k_B T \int^\Lambda {d^2p \over (2\pi)^2} \sum_{n=1}^N \ln \left[ 1 -
\exp(-\varepsilon_{{\bf p},n}/(k_B T)) \right],
\end{equation}
where $k_B$ and $T$ denote the Boltzmann constant and the temperature,
respectively. An additional factor of two coming from the summation over the
polarizations has already been included in Eq.(\ref{F}). The integration over
${\bf p}$ is carried out to an ultraviolett cutoff $\Lambda$ and the
sum is truncated at some maximum mode number $N$. The ultraviolett
cutoff parameter $\Lambda$ is typically determined by the radius of the
first Brillouin zone of the plate material. If $a$ is the lattice constant
of the material we identify $\Lambda = \pi/a$. The maximum mode number $N$
can be written as $N = L / b$, where $b$ is also a microscopic length scale
(see below).

For $T = 0$ only the first term in Eq.(\ref{F}) remains and this is the first
example studied by Casimir \cite{Casimir}. Here, we will
discuss the high temperature limit $k_B T \gg \hbar c \Lambda$, because
this allows us to illustrate some aspects of the calculations involved
within continuum models like, e.g., the Ginzburg - Landau model on a rather
elementary level. For the more general case of layered dielectrica at finite
temperature (dispersion forces) we refer to the classical literature
\cite{Dispersion,Israel} and to Ref.\cite{Reviews}. To leading order in
$\hbar c \Lambda / (k_B T)$ we obtain
\begin{equation}
\label{FhighT}
F = 2 k_B T \int^\Lambda {d^2p \over (2\pi)^2} \sum_{n=1}^N \ln
{\varepsilon_{{\bf p},n} \over k_B T} .
\end{equation}
Note that the integral and the summation in Eq.(\ref{FhighT}) are only
meaningful for finite cutoff parameters $\Lambda$ and $N$. However, we
only need the final result in the limits $\Lambda L \gg 1$ and $N \gg 1$.
For simplicity we identify $k_B T \sim \hbar c \pi / b$ to the order of
magnitude which implies $\Lambda L \ll N$, i.e., $b \ll a$. The integral in
Eq.(\ref{FhighT}) is elementary and the resulting terms can be arranged as
\begin{eqnarray}
\label{FhTsum}
F &=& {k_B T \over 4 \pi} \Lambda^2 \sum_{n=1}^N \left\{ 2 \ln\left(
{\hbar c \over k_B T}\ {n \pi \over L}\right) \right.
\\ \nonumber \\
&+& \left. \ln\left( 1 + \left( {\Lambda L \over n \pi} \right)^2 \right) +
\left({n \pi \over \Lambda L}\right)^2 \left[
\ln\left( 1 + \left( {\Lambda L \over n \pi} \right)^2 \right) -
\left( {\Lambda L \over n \pi} \right)^2 \right] \right\}.
\nonumber
\end{eqnarray}
The sum over the terms in the second line of Eq.(\ref{FhTsum}) converges
so that we can immediately perform the limit $N \to \infty$ here. We obtain
\begin{equation}
\label{FhT}
F = {k_B T \over 4 \pi} \Lambda^2 \left\{ 2 \ln N! +
2N \ln \left( {\hbar c \over k_B T}\ {\pi \over L}\right) +
\int_0^1 \ln {\sinh(\Lambda L \sqrt{x}) \over \Lambda L \sqrt{x}} dx
\right\},
\end{equation}
where terms which {\em vanish} in the limit $N \to \infty$ have already been
dropped. In order to evaluate Eq.(\ref{FhT}) further for large $N$ and
$\Lambda L$ we employ Stirlings formula and the series expansion of the
logarithm. With $\Lambda = \pi / a$ and $N = L / b$ we obtain the final result
\begin{equation}
\label{FhTfin}
F = L\ {\pi k_B T \over 2 a^2 b} \left[ \ln \left(
{\hbar c \over k_B T}\ {\pi \over b}\right) - 1 + {\pi b \over 3 a}\right]
+ {\pi k_B T \over 8 a^2}\left[2 \ln \left({a \over b}\right) + 1 \right]
- {k_B T \over L^2}\ {\zeta(3) \over 8 \pi} + \dots
\end{equation}
where $\zeta(3) \simeq 1.202$ is a special value of the Riemann zeta function
and the dots indicate contributions which are exponentially small in $\Lambda
L$. All terms which vanish in the limit $N \to \infty$ ($b \to 0$) have
consistently been dropped.

The decomposition of the free energy per unit area given by Eq.(\ref{FhTfin})
is a special case of the general decomposition
\begin{equation}
\label{Fdecomp}
F = L F_b + F_{s,a} + F_{s,b} + \delta F_{ab}
\end{equation}
for a film with two surfaces of type $a$ and $b$. The leading contribution
to $F$ is proportional to $L$ and it corresponds to the {\em bulk}
contribution of the free energy. In our example it is given by the radiation
pressure $F_b$ between the plates. The second contribution to Eq.(\ref{FhTfin})
is independent of $L$ and it therefore corresponds to the sum of the surface
free energies or {\em surface tensions} $F_{s,a} + F_{s,b}$, where $a = b$
in the above example. The third contribution varies as $L^{-2}$ and it
corresponds to the fluctuation induced long - ranged {\em Casimir interaction}
between the plates, which is the most prominent contribution to the
{\em finite-size} part $\delta F_{ab}$ of the free energy
in our example. Note that the Casimir contribution is {\em
independent} of the microscopic cutoff parameters $a$ and $b$. Its absolute
strength at a given distance $L$ and temperature $T$ is governed by the
numerical constant $\Delta = -\zeta(3) / (8\pi) \simeq -0.0478$ which is
usually called the {\em Casimir amplitude}. The Casimir amplitude is negative
here, so that the Casimir force between the plates is {\em attractive}. The
Casimir interaction can be obtained in very elegant ways known as zeta -
function regularization, algebraic cutoff, or exponential cutoff schemes
\cite{Elizalde}. Their equivalence with respect to the cutoff - independent
Casimir interaction has been explicitly shown for the example presented here
\cite{Svaiter}. For further details see also Ref.\cite{Elizalde94}.

\subsection{Critical phenomena and correlated fluids}
The above example for the Casimir effect appears to be very specific at
first sight, but the functional form of the free energy given by
Eqs.(\ref{FhighT}) and (\ref{FhTfin}) is far more general than it seems.
In fact, the underlying mechanism which leads to fluctuation induced
long - ranged forces only requires a fluctuating field with geometric
restrictions and a macroscopic length scale $L$ imposed by the geometry as
the only limiting factor for the wavelength of the fluctuations. Any system
which is at a {\em critical point} also meets this requirement. The
fluctuating field in this case is given by the order parameter and each of
the individual contributions to the free energy as given by Eq.(\ref{Fdecomp})
is a sum of a {\em regular} part and {\em singular} part which contains the
critical behavior of the system. Right {\em at} the critical point the
correlation length $\xi$ is infinite so that the distance $L$ between the
system boundaries provides the only macroscopic length scale as required for
the occurrence of long - ranged fluctuation induced forces. Above the critical
point the correlation length is finite and therefore the ratio $\xi / L$
governs the range of these forces. The existence of Casimir forces in critical
systems was anticipated by Fisher and de Gennes \cite{FdG78} in the framework
of the so-called distant wall corrections to critical profiles, where in
many cases the Casimir amplitudes govern the leading distant-wall correction
term to the profile in the vicinity of one of the system boundaries. For
details and an extendend list of further references the reader is referred
to chapter 4 of Ref.\cite{MKbook} and to Ref.\cite{EEMKSD96}.

It is important to realize that the Casimir amplitudes $\Delta$ and the
associated scaling functions $\theta(L/\xi)$ that take the place of these
amplitudes for {\em finite} $L / \xi$ (see Ref.\cite{MKbook} and Sec.2)
are {\em universal}, i.e., they do not depend on microscopic details of the
system under consideration. Note, however, that the precise form of the
scaling functions $\theta$ depends on the {\em definition} of the correlation
length $\xi$. For systems with surfaces the concept of universality classes
raises the question of whether there is {\em surface} critical behavior and
to what extent it is governed by {\em universal surface critical exponents}.
During the 1980's this question was answered in favor of the general ideas
of critical behavior and universality, i.e., microscopic surface properties
are indeed unimportant. One only has to specify the type of {\em boundary
condition} which the surface imposes on the order parameter. In this respect
there are only three fundamentally different {\em surface universality classes}
\cite{SurfRev}. In particular, the surface may enhance the order parameter
with the result that the system undergoes a second order phase transition
in presence of an already {\em ordered} surface (extraordinary transition, E).
The surface may also suppress the order parameter with the result that the
system undergoes a second order phase transition in presence of a {\em
disordered} surface (ordinary transition, O). Finally the surface and the
bulk may order at the {\em same} temperature (special transition, SB), so
the critical point of the system is in fact a {\em multicritical} point. If
the spatial dimensionality $d$ of the system is high enough there are two
options for the occurrence of surface order above the bulk critical
temperature $T_{c,b}$. The surface may order {\em spontaneously} at a
certain critical temperature $T_{c,s} > T_{c,b}$ {\em or} the surface may
be ordered {\em externally} by the presence of a {\em surface field}. The
bulk transition in presence of an externally ordered surface is called
{\em normal} transition. However, it has been shown recently by rigorous
arguments that the normal and the extraordinary transitions only differ
by corrections to scaling, so both belong to the extraordinary
surface universality class \cite{BD94}. Surface critical behavior has
already been extensively reviewed \cite{SurfRev} (see also chapter 2 of
Ref.\cite{MKbook} for a short summary), so we refrain from giving
further details here.

The distinction between the surface universality classes is vital for the
Casimir forces, because the Casimir amplitudes $\Delta$ and the scaling
functions $\theta$ depend on them. The simplest boundary conditions apart
from periodic ones are {\em Dirichlet} boundary conditions which suppress
the order parameter to zero at the surface. A system with these boundary
conditions provides a representation of the ordinary surface universality
class. The first systematic field-theoretic calculation of a Casimir
amplitude was done by Symanzik \cite{Sym81} for this case. Starting from
Eq.(\ref{FhighT}) the above example essentially reproduces all neccessary
steps for such a calculation at the one - loop level (Gaussian theory).
In general, concepts of the field - theoretic renormalization group are
required which cannot be described here. The application of field theory
to the critical behavior of finite systems is a field of ongoing research
\cite{FTFSS} which has recently furnished unexpected results concerning
the occurrence of {\em nonuniversal} critical finite - size behavior above
the upper critical dimension \cite{CD98}. For reviews about the general
concept of critical finite - size scaling the reader is referred to
Ref.\cite{FSSRev}.

Above the critical temperature the range of the Casimir force is always
limited by the correlation length, but below the critical temperature the
situation may be different. In Ising - like systems the correlation length
is also finite below $T_{c,b}$ and therefore the Casimir forces have a
finite range. If the system has a continuous symmetry, however, {\em
Goldstone modes} cause correlation functions of the order parameter to remain
long - ranged below $T_{c,b}$. The most prominent
examples are XY and Heisenberg ferromagnets which posess an $O(N)$
symmetry with $N = 2$ and $N = 3$, respectively, in contrast to the Ising
ferromagnet $(N = 1)$. In other words, the correlation length of continuous
ferromagnets remains infinite below $T_{c,b}$ and therefore Goldstone
modes also give rise to fluctuation induced long - ranged forces between
system boundaries. Fluids with this property are sometimes called {\em
correlated fluids}. The most important examples with respect to experimental
realizations are liquid $^4$He below the superfluid - normal transition
\cite{Li92} and nematic liquid crystals in the nematically ordered phase
\cite{APP91}, where fluctuations of the nematic director are responsible
for the long - ranged nature of the Casimir force. Near the phase transition
to the isotropic phase fluctuations of the degree of nematic order and the
degree of biaxiality generate {\em short - ranged} corrections to the Casimir
force \cite{Ziherl98}.

In summary, we have mentioned three options for the occurrence of
fluctuation - induced long - ranged forces: the presence of long - ranged
interactions (e.g., electromagnetism, see Sec.1.A), the presence of critical
fluctuations, and the presence of Goldstone modes. In the following overview
we will only consider the second option, i.e., systems in the vicinity of
critical points. In particular, recent progress in the theoretical
understanding of critical Casimir forces for all surface universality classes
and especially for curved geometries will be presented.
Special attention is also paid to the comparison of Casimir amplitudes and
corresponding scaling functions with computer simulations and experiments,
which are still in progress at this time. Due to the limited scope of this
article other interesting developments in related areas will not be described
in any detail and an apology is made in advance to all authors whose
work is not explicitly mentioned here. The remainder of this article is
organized according to the three main approaches to critical Casimir forces,
namely analytic theory, computer simulation, and experiments.

\section{Analytic theory}
The analytic theory of the Casimir effect in critical systems is based on
the concept of finite - size scaling \cite{MKbook,FSSRev}. Exact solutions
of model systems in statistical mechanics give only limited access to
the finite - size scaling functions, because they are mainly restricted to
two dimensional systems. In $d \geq 3$ dimensions only the spherical model
can be analyzed in a rigorous fashion \cite{Spherical} which has recently
been done with special regard to the film geometry for $d = 3$ dimensions
\cite{Danchev96,Danchev98}. Despite their limitations exact solutions provide
valuable insight into the structure of the scaling functions
and sometimes the results for $d = 2$ can be used to improve estimates
obtained by approximative methods for $d = 3$ (see Sec.3 of Ref.\cite{MKbook}
and below).

The concept of finite - size scaling is a natural extension of the principle
of {\em scale invariance} to critical systems with geometric constraints on
macroscopic length scales. The principle of scale invariance itself may
be viewed as a special case of the more general principle of {\em conformal
invariance} (see Sec.3 of Ref.\cite{MKbook} and Ref.\cite{ConfInv}). Conformal
invariance implies the equivalence of systems with boundaries at $T = T_{c,b}$
if these systems can be mapped onto one another by a conformal transformation.
The principle of conformal invariance holds in any dimension, but it is
particularly powerful for $d = 2$ due to the exceptional large number
of conformal mappings in this case (large conformal group, see
Ref.\cite{ConfInv}). Note that scale transformations are just very special
conformal mappings. In the framework of conformal field theory the {\em
stress tensor} plays a key role \cite{ConfInv,MO93}. Here we only mention that
the thermal average of the stress tensor yields the local Casimir force in
a critical system and therefore the stress tensor provides a very important
tool in the analytic theory of the Casimir effect. In fact, most of the
Casimir amplitudes for $d = 2$ have been obtained from conformal field
theory rather than exact solutions (see Sec.3 of Ref.\cite{MKbook} and
Ref.\cite{ConfInv}).

Many of the experimentally relevant results have been obtained from
a field - theoretic analysis of the well - known Ginzburg - Landau Hamiltonian
$\cal H$ with geometric constraints which can be decomposed according to
${\cal H} = {\cal H}_b + {\cal H}_s + {\cal H}_e$. The bulk contribution
${\cal H}_b$ is given by
\begin{equation}
\label{Hb}
{\cal H}_b = \int_V d^dx \left[ {1 \over 2} \left(\nabla \Phi \right)^2 +
{\tau \over 2} \Phi^2 + {u \over 4!} \left(\Phi^2\right)^2 -
{\bf H} \cdot \Phi \right]
\end{equation}
for systems characterized by a $N$-component order parameter $\Phi =
(\phi_1, \dots , \phi_N)$ confined to a volume $V$, where $N = 1,2,3$
characterize the Ising, XY, and Heisenberg universality class, respectively.
The parameters $\tau$ and $\bf H$ correspond to the bare reduced temperature
and external field. The physical (renormalized) reduced
temperature and external field will be denoted by $t$ and $h$ in the following.
The surface contribution ${\cal H}_s$ can be written as
\begin{equation}
\label{Hs}
{\cal H}_s = \int_S d^{d-1}x \left[ {c \over 2} \Phi^2 -
{\bf H}_1 \cdot \Phi \right],
\end{equation}
where $c$ and ${\bf H}_1$ correspond to the surface enhancement and the
surface field, respectively \cite{SurfRev}. Note, that the surface $S$ may
consist of serveral disjoint parts. The last contribution ${\cal H}_e$
contains edge and curvature contributions to the Hamiltonian $\cal H$ which
have first been considered in Ref.\cite{Sym81} within the framework of the
renormalization group. For experiments the ordinary transition $(c = \infty)$
and the extraordinary transition (e.g., $c = -\infty$, see also
Ref.\cite{BD94}) are the most important cases. Neither ${\cal H}_s$ nor
${\cal H}_e$ needs to be considered here in any detail, because their effect
is completely contained in the boundary conditions for the order parameter
$\Phi$. We will therefore restrict the following discussion to the ordinary
and the extraordinary surface universality class and to periodic boundary
conditions.

\subsection{The spherical model}
The spherical model can be considered as the $N \to \infty$ limit of $O(N)$
symmetric classical spin models and it can also be expressed as the $N \to
\infty$ limit of Eqs.(\ref{Hb}) and (\ref{Hs}). We only summarize the most
recent results here, for a brief overview the reader is referred to Sec.2.2
of Ref.\cite{MKbook} and to Ref.\cite{Spherical}. In the presence
of an external field $h$ and for sufficiently small values of the reduced
temperature $t = (T - T_{c,b})/T_{c,b}$ the {\em singular} contribution
$\delta f_{ab}$ to the finite-size part $\delta F_{ab}$ of the free energy
per unit area in a film geometry (see Eq.(\ref{Fdecomp})) in $d$
dimensions can be cast into the scaling form \cite{Danchev96}
\begin{equation}
\label{df}
\delta f_{ab}(t, h, L) = k_B T_{c,b}L^{-(d-1)} \theta_{ab}
\left( t L^{1/\nu}, h L^{\beta \delta / \nu} \right)
\end{equation}
near the critical point given by $t = 0$ and $h = 0$,
where $ab$ indicates the combination of surface universality classes at the
two surfaces. The critical exponents $\nu$ and $\beta$ characterize the
temperature dependence of the correlation length $\xi \sim t^{-\nu}$, $t>0$
and the order parameter (spontaneous magnetization) $m \sim (-t)^\beta$,
$t<0$ for $h=0$, respectively. The exponent $\delta$ characterizes the
functional dependence of the magnetization $m \sim |h|^{1/\delta}$ on the
external field $h$ for $t=0$. The form of the scaling arguments in
Eq.(\ref{df}) is imposed by the principle of scale invariance. They can be
obtained by observing that $L/\xi$ is equivalent to the first scaling
argument and $L/\xi_h$, where $\xi_h \sim h^{-\nu/(\beta \delta)}$ is the
correlation length for finite field $h$ at $t=0$, is equivalent to the
second scaling argument. For nearest neighbor interactions for $d = 3$
the critical exponents $\nu$, $\beta$, and $\delta$ of the spherical model
are given by $\nu = 1$, $\beta = 1/2$, and $\delta = 5$. The special value
$\Delta_{ab} \equiv \theta_{ab}(0,0)$ of the scaling function is the Casimir
amplitude. In units of $k_B T_{c,b}$ the Casimir force ${\cal K}_{ab} \equiv
-{\partial \over \partial L} \delta f_{ab}$ is characterized by the
corresponding scaling form
\begin{equation}
\label{K}
{\cal K}_{ab}(t, h, L) = L^{-d} K_{ab} \left( t L^{1/\nu},
h L^{\beta \delta / \nu} \right),
\end{equation}
where the scaling function $K_{ab}$ is given by
\begin{equation}
\label{Kab}
K_{ab}(x,y) = (d - 1)\theta_{ab}(x,y)
- {1 \over \nu} x {\partial \over \partial x} \theta_{ab}(x,y)
- {\beta \delta \over \nu} y {\partial \over \partial y} \theta_{ab}(x,y).
\end{equation}
The universal scaling functions $\theta_{ab}$ and $K_{ab}$ have been
investigated recently for periodic boundary conditions $ab = per$
\cite{Danchev96,Danchev98}. For $h = 0$ the scaling functions
$\theta_{per}(x,0)$ and $K_{per}(x,0)$ are both negative and increase
monotonically with x, i.e., unlike the scaling functions in Ising - like
systems they do not have a minimum in the vicinity of $T = T_{c,b}$ $(x = 0)$
\cite{Danchev96}. For $x \to +\infty$ both scaling functions decay to zero
exponentially, whereas for $x \to -\infty$ $K_{per}(x,0) \to -\zeta(3)/\pi
\simeq -0.382$. This behavior is due to the presence of {\em Goldstone modes}
in the spherical model below $T_{c,b}$. For finite values of $h$ $(y \neq 0)$
the scaling functions $\theta_{per}(x,y)$ and $K_{per}(x,y)$ again decay
exponentially as $L \to \infty$ \cite{Danchev96}. Moreover, both scaling
functions also decay exponentially for $y \to \infty$ at $x = 0$. The
Casimir amplitude
\begin{equation}
\label{Dpersph}
\Delta_{per} = \theta_{per}(0,0) = -{2 \zeta(3) \over 5 \pi} = -0.15305\dots
\end{equation}
can be obtained exactly \cite{Danchev98} and numerically it is very close the
best availabe estimates for the Ising model for $d = 3$ (see Table
\ref{Dtable} in Sec.3). It has also been shown rigorously, that the scaling
function $\theta_{per}(x,y)$ is a monotonically increasing function of each
of its arguments as long as the temperature $T$ is in the vicinity of
$T_{c,b}$ \cite{Danchev98}. However, the {\em hypothesis} that this statement
is true for {\em any} nearest neighbor $O(N)$ symmetric spin model for $N
\geq 2$ \cite{Danchev98} could not be substantiated so far (see below).

\subsection{The Ginzburg Landau model}
\subsubsection{Film geometry}
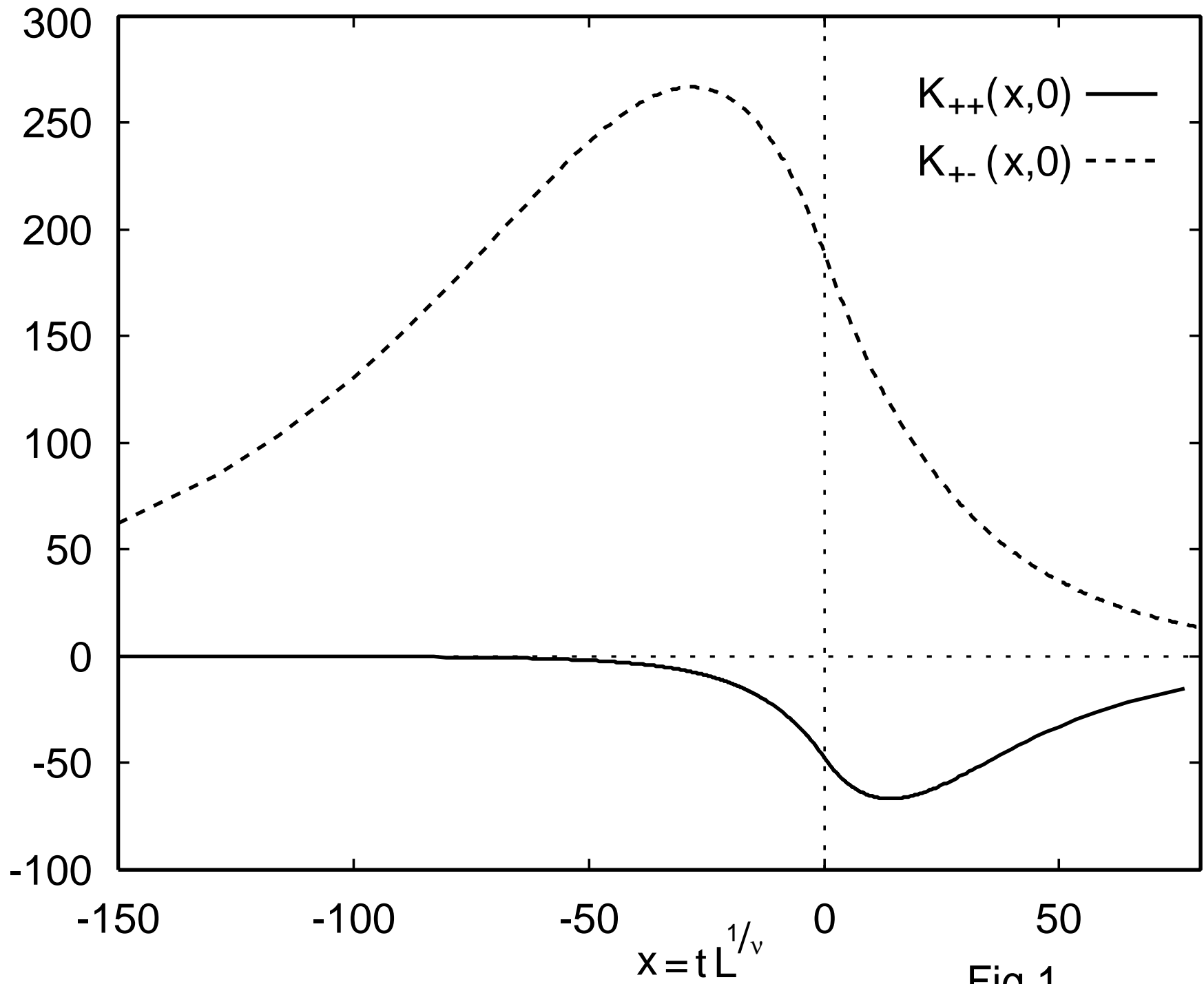
\begin{figure}[t]
\caption{Scaling functions $K_{++}(x,0)$ (solid line) and $K_{+-}(x,0)$
(dashed line) taken from Fig.1 of Ref.\protect\cite{MK97}. The $x$ range
influenced by the bulk critical point $x = 0$ is very broad and the asymptotic
decay for $x \to \pm \infty$ is dominated by an exponential. Note that
$K_{++}(x,0)$ and $K_{+-}(x,0)$ take their extreme values at $x \simeq 10$
and $x \simeq -25$, respectively.
\label{Kx0}}
\end{figure}

The film geometry has also been reinvestigated for the Ginzburg Landau model
for the case of the extraordinary surface universality class \cite{MK97},
which is of particular interest for experiments with critical binary liquid
mixtures. The scaling functions $K_{ab}(x,0)$ in zero external field have
been determined within mean field theory for infinitely strong surface fields
${\bf h}_1$ and ${\bf h}_2$ which enclose an arbitrary angle $\alpha$ between
$\alpha = 0$ (parallel surface fields) and $\alpha = \pi$ (antiparallel
surface fields). For Ising-like systems only $\alpha = 0$ and $\alpha = \pi$
can be realized and we refer to these cases as $ab = ++$ and $ab = +-$.
The Casimir amplitude is negative for $\alpha = 0$ and positive for
$\alpha = \pi$, it changes sign at $\alpha = \pi/3$ \cite{MK97}. Accordingly,
the scaling function $K_{++}(x,0)$ is negative and the scaling function
$K_{+-}(x,0)$ is positive for all $x$ within mean field theory, but it seems
very likely that this is also true beyond mean field theory. The functional
form of $K_{++}(x,0)$ and $K_{+-}(x,0)$ is illustrated in Fig.\ref{Kx0},
where the normalization of Ref.\cite{MK97} has been used. Note that both
scaling functions take their extremal values at nonzero $x$, which makes
them qualitatively very similar to the corresponding scaling functions for
an Ising strip for $d = 2$, which can be solved exactly \cite{ES94}.

The one-loop corrections to the mean field behavior are very hard to otain
and at present they only exist for the Casimir amplitudes in the form of the
$\varepsilon$-expansion, where $\varepsilon = 4 - d$. The numerical quality
of the $\varepsilon$-expansion when extrapolated to $\varepsilon = 1$
is very poor so that exact results for $d = 2$ have been included in order
to obtain an interpolation formula for the Casimir amplitudes for $d = 3$.
The values for $\Delta_{++}$, $\Delta_{+-}$, and
$\Delta_{O+}$ obtained in this way agree reasonably well with Monte-Carlo
estimates (see Sec. 3) \cite{MK97}. The $\varepsilon$-expansion
for $\Delta_{++}$ and $\Delta_{+-}$ has recently received some independent
confirmation from local-functional methods \cite{BU98} which also provide
reliable numerical estimates for $d = 3$ (see Table \ref{Dtable} in Sec. 3).

Apart from usual critical points, for which the upper critical dimension is
$d_u = 4$, tricritical points in liquid mixtures with more than two components
\cite{URMG97} and in $^3$He - $^4$He mixtures (see Sec.6 of Ref.\cite{MKbook})
also provide an opportunity for experimental tests of the Casimir force.
A theoretically appealing feature of a tricritical point is, that its upper
critical dimension is $d_u = 3$ so that exact results for $d = 3$ can be
obtained essentially from a mean field or a Gaussian theory. If, for example,
Dirichlet boundary conditions are the correct ones for a $^3$He - $^4$He
mixture in a film at the tricritical point, then the Casimir amplitude
$\Delta_{OO} = -\zeta(3)/(8\pi) \simeq -0.0478$ given in Eq.(\ref{FhTfin})
is also the right one for this system. There is, however, some debate
concerning the correct boundary conditions for tricritical $^3$He - $^4$He
mixtures \cite{URMG97}. The result obtained for $\Delta_{++}$ at a
tricritical point for $d = 3$ contains a logarithmic factor which is absent
below the upper critical dimension $d_u$ and which introduces a dependence
on a microcopic length scale into the Casimir amplitude \cite{URMG97}.
This depencence is very weak and $\Delta_{++}$ at tricriticality is expected
to be about seven times larger than the corresponding amplitude at a usual
critical point \cite{URMG97}.

\subsubsection{Curved geometries}
In view of experimental set-ups for, e.g., atomic force microscopy it is
desirable to consider other geometries than films, because two plates cannot
be kept parrallel accurately enough during force measurements. Curved
geometries like a sphere in front of a planar wall or two spheres are much
more convenient to control experimentally and are also much closer to
reality in, e.g., colloidal suspensions \cite{DBDE85} (see also
Ref.\cite{SCP}). Some theoretical effort has therefore been made on the
investigation of these curved geometries, where conformal invariance
considerations have proved to be a very powerful tool at the critical point
\cite{TBEE95}. If ${\cal F}_{ab}(r,R_1,R_2)$ denotes the free energy of a
critical fluid in which two spheres with radii $R_1$ and $R_2$ at a
center - to - center distance $r > R_1 + R_2$ are immersed, then the Casimir
interaction $\delta {\cal F}_{ab}$ takes the scaling form \cite{TBEE95}
\begin{equation}
\label{dFab}
\delta {\cal F}_{ab}(r,R_1,R_2) \equiv {\cal F}_{ab}(r,R_1,R_2) -
{\cal F}_{ab}(r = \infty,R_1,R_2) = k_B T_{c,b} F_{ab}(\kappa),
\end{equation}
where $ab$ denotes the combination of surface universality classes and
$\kappa$ is the conformally invariant cross ratio
\begin{equation}
\label{kappa}
\kappa = (2 R_1 R_2)^{-1} |r^2 - R_1^2 - R_2^2|.
\end{equation}
Note that the cases of two separate spheres in an unbounded critical medium
and a single sphere inside a critical medium of spherical shape are conformally
equivalent and are therefore governed by the same universal scaling function
$F_{ab}(\kappa)$ \cite{TBEE95}. For large $R_1$ and $R_2$ at fixed surface -
to - surface distance $D = r - R_1 - R_2$ one obtains from the limit of
parallel plates \cite{TBEE95}
\begin{equation}
\label{dFabpp}
\delta {\cal F}_{ab}(r,R_1,R_2) = k_B T_{c,b} S_d \Delta_{ab}
\left[2 \left(D/R_1 + D/R_2\right) \right]^{-(d-1)/2} ,
\end{equation}
where $S_d$ is the surface area of the unit sphere in $d$ dimensions and
$\Delta_{ab}$ is the Casimir amplitude for parallel plates. In the opposite
limit $r \gg R_1, R_2$ the presence of the spheres can be taken into account
by the small sphere expansion \cite{TBEE95} which yields
\begin{equation}
\label{dFabss}
\delta {\cal F}_{ab}(r,R_1,R_2) = -k_B T_{c,b} {A_a^{\psi} A_b^{\psi} \over
B_\psi} \left({R_1 R_2 \over r^2}\right)^{x_\psi},
\end{equation}
where $\psi = \phi$ is the order parameter if both $a = b = E$ indicate the
extraordinary surface universality class. In this case the scaling exponent
$x_\psi$ is the scaling exponent of the order parameter $x_\phi = \beta / \nu$
($\simeq 0.517$ for the Ising model for $d = 3$). If only one of the surfaces
is not characterized by the extraordinary surface universality class, the
operator $\psi$ is given by the local energy density $\phi^2$ and $x_\psi$
is the corresponding scaling exponent $x_{\phi^2} = d - 1/\nu$ ($\simeq 1.41$
for the Ising model for $d = 3$). The amplitudes $A_a^{\psi}$ and $A_b^{\psi}$
are the amplitudes of the critical profiles $\langle \psi(z)
\rangle^s_{\infty/2} = A_s^{\psi} (2z)^{-x_{\psi}}$, $s = a,b$ of the operator
$\psi$ in a seminifinite system bounded by a planar surface of type $s$.
The amplitude $B_\psi$ is the amplitude of the $\psi \psi$ correlation
function in unbounded space. Although none of these amplitudes is universal
individually, their combination in Eq.(\ref{dFabss}) is universal and its
value for various surface types is exactly known for the Ising universality
class for $d = 2$. In $d = 4 - \varepsilon$ dimensions estimates can be
calculated from a renormalization group analysis of the Ginzburg Landau model
\cite{TBEE95}. Note, that Eqs.(\ref{dFab}), (\ref{dFabpp}), and (\ref{dFabss})
only hold at the critical point. The Casimir interaction according to
Eq.(\ref{dFabss}) is in fact very long ranged. For the extraordinary surface
universality class it decays about as slowly as the Coulomb interaction.
In all other cases the decay is faster, but it is still slower than
the decay of, e.g.,  dipolar interactions.

The full functional form
of the scaling funcions $F_{++}(\kappa)$, $F_{+-}(\kappa)$, $F_{+SB}(\kappa)$,
and $F_{+O}(\kappa)$ has been calculated within mean field theory from the
stress tensor in the concentric sphere geometry \cite{EEUR95}. As for the case
of parallel plates $F_{++}$ and $F_{+SB}$ are negative (attractive Casimir
force), whereas $F_{+-}$ and $F_{+O}$ are positive (repulsive Casimir force).
The boundary conditions $ab = OO$, $O SB$, and $SB SB$ have been treated
within the Gaussian model, where $F_{OO}(\kappa) = F_{SB\ SB}(\kappa) < 0$
and $F_{O\ SB}(\kappa) > 0$ has been found. Although the analytic information
from mean field or Gaussian theory is quite limited, the combination of these
results with exact results for $d = 2$ yields fairly reliable estimates
for $ab = ++$, $+-$, $+O$, and $OO$ within the Ising universality class in
$d = 3$ \cite{EEUR95}. Higher order calculations beyond mean field or the
Gaussian approximation, respectively, for the concentric geometry are
extremely demanding and results are not available. Finally, we note that
the sphere - planar wall (SPW) geometry can also be obtained from the
concentric geometry by a conformal mapping \cite{EEUR95}.

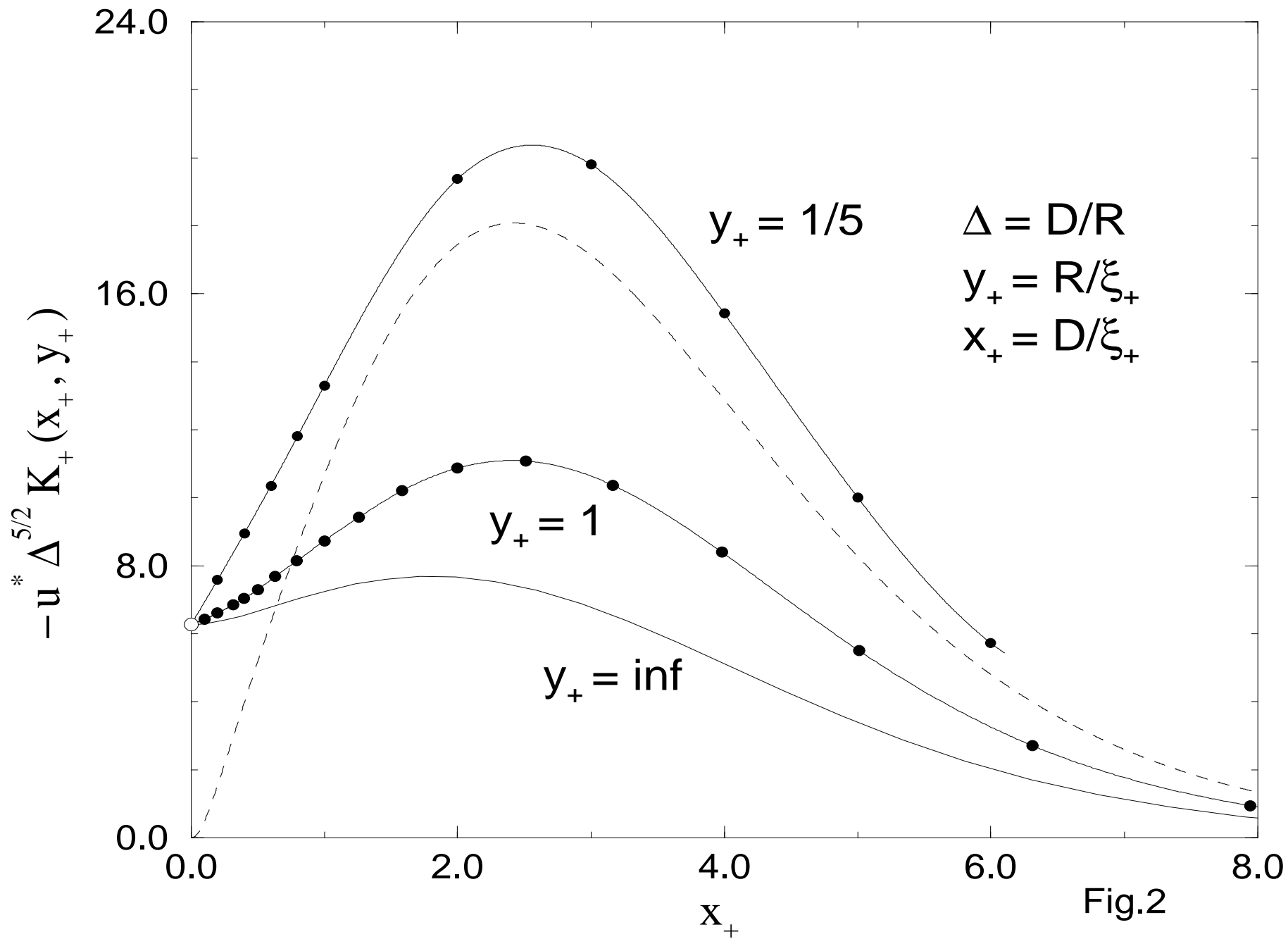
\begin{figure}[t]
\caption{Scaling function $K_{++}^+(x_+,y_+)$ as function of $x_+$ for
various values of $y_+$ (solid lines). The prefactor $\Delta^{5/2}$ absorbs
the divergence of the scaling function in the limit $\Delta \to 0$, where
the Derjaguin approximation becomes valid (see main text). The fixed point
value $u^*$ of the renormalized coupling constant is required as an additional
normalization due to the mean field character of the calculation. The dashed
line corresponds to the small sphere expansion, which is shown here for
$y_+ = 1/5$. The exponential decay of the scaling function sets in at
$x_+ \simeq 4$.
\label{KSPWplot}}
\end{figure}

So far conformal invariance could be used to obtain the scaling functions
of the Casimir interaction for various geometries with spherical surfaces.
If the correlation length $\xi$ is finite, conformal invariance does
no longer hold. Moreover, if all length scales $\xi$, $r$, $R_1$, and $R_2$
are comparable, small sphere expansions cannot be made any longer and a new
calculation is required for every geometry. In this case even mean
field results can only be obtained numerically \cite{HSED98}. So far, this
has only been done in detail for the SPW geometry $(R_1 = R, R_2 \to \infty,
D = r - R_1 - R_2 = const.)$ with $++$ boundary conditions and at arbitrary
temperature near the critical point for Ising - like systems \cite{HSED98}.
We restrict the discussion to the case $T > T_{c,b}$, where the correlation
length $\xi_+ = \xi^0_+ t^{-\nu}$ governs the decay of the order parameter
correlation function in real space. The Casimir force can be cast into the
scaling form \cite{HSED98}
\begin{equation}
\label{KSPW}
{\cal K}_{++}(t,D, R) = {k_B T_{c,b} \over R} K_{++}^+
\left(x_+ = D/\xi_+, y_+ = R/\xi_+\right),
\end{equation}
where a corresponding scaling function $K_{++}^-(x_-,y_-)$ governs the scaling
behavior of the Casimir force below $T_{c,b}$. The scaling functions are
obtained from the mean field evaluation of the stress tensor which requires
the knowledge of the order parameter profile within mean field theory. The
order parameter profile is obtained from a numerical solution of the Euler -
Lagrange equation for Eq.(\ref{Hb}) in presence of parallel and infinite
surface fields which dictate the boundary conditions. The functional form
of $K_{++}^+(x_+,y_+)$ is illustrated in Fig.\ref{KSPWplot}. As for the case
of parallel plates, the Casimir force is attractive and takes its maximum
value above $T_{c,b}$. The true position of the maximum is somewhat concealed
in Fig.\ref{KSPWplot} due to the normalization factor $\Delta^{5/2}$, which is
required in order to absorb the divergence of $K_{++}^+$ for $\Delta = D/R
\to 0$. In this limit the Derjaguin approximation becomes valid, where the
Casimir force is represented as an integral over parallel plate contributions.
Each of these ``parallel plates'' in the $d$ -dimensional SPW geometry is an
infinitesimal annulus of width $d\rho$ and radius $\rho$ which is located on
the surface of a paraboloid in order to approximate the sphere near the wall.
The distance of one of these annuli from the wall is then given by $L(\rho) =
D + \rho^2/(2R)$, where the integration is performed from $\rho = 0$ to $\rho
= \infty$ \cite{HSED98}. Note that this approximation is only valid for
forces which decay sufficiently fast as $L(\rho) \to \infty$. The
amplitude of the Derjaguin approximation to the Casimir force at $T = T_{c,b}$
is indicated by the open circle in Fig.\ref{KSPWplot}, where all solid lines
meet. The dashed line correponds to the small sphere expansion to leading
order, where $y_+ = 1/5$ has been used instead of the correct choice $y_+ = 0$,
for which a factor $\Delta^2$ is required for proper normalization
\cite{HSED98}.

The presence of small external fields can be used to drive, e.g., a critical
binary liquid mixture slightly away from the critical concentration. Within
the small sphere expansion the Casimir energy between two spheres (colloidal
particles) turns out to be nonsymmetric with respect to deviations from the
critical concentration so that the Casimir force is enhanced when the
concentration of the component preferentially adsorbed by the colloids
is reduced \cite{HSED98}. This asymmetry is consistent with the asymmetry
found experimentally in the flocculation phase diagram of colloidal
suspensions \cite{DBDE85}.

The concentric sphere geometry for $++$ boundary conditions has also been
considered at tricritical points \cite{URMG97}, where the principle of
conformal invariance can be used as well (see Ref.\cite{EEUR95}). The
expressions for the Casimir forces in the different geometries are similar
to the ones obtained for critical points \cite{EEUR95} apart from the
logarithmic dependence on a microscopic length scale. The scaling function
of the Casimir force depends on the conformally invariant cross ratio given
by Eq.(\ref{kappa}). In the range of distances $D$, where force measurements
with the atomic force microscope appear to be feasable, both Casimir and van
der Waals forces are essentially governed by the parallel plate limit of the
curved geometries studied here \cite{URMG97}. Corresponding investigations
of the Casimir forces away from the tricritical point have apparently not
been performed.

Finally, we mention that a diluted polymer solution may also serve as a
critical medium which mediates long - ranged forces between, e.g., colloidal
particles \cite{EHD96}. Systematic investigations, however, are still at an
early stage and the description of these is beyond the scope of this article.

\section{Computer simulation}
Computer simulations of forces in liquid films have been performed in the
past primarily with regard to the microscopic mechanisms of friction,
adhesion, and lubrication, where both Monte - Carlo \cite{SDC94} and
molecular dynamics methods \cite{DSS96} have been used (see Ref.\cite{Israel}
for background material and more details). With regard to the Casimir force
in critical or correlated fluids the situation is less satisfactory. The
computational effort involved in such calculations is substantial and
consequently only very few Monte - Carlo studies of the critical Casimir
effect exist. Only rectangular geometries have been considered so far in
$d = 3$, because the currently available system sizes do not provide
sufficient resolution to investigate curved geometries.

\subsection{Casimir amplitudes}
The first systematic
attempt to measure the Casimir amplitudes of Ising and Potts models in a
film geometry is based on a splitting procedure for lattice models {\em at}
criticality \cite{KL96}. The systems contain $M^{d-1} \times L$ lattice sites,
where an aspect ratio of $M/L = 6$ turns out to be sufficient to approximate
the film geometry. In the lateral directions always boundary conditions
are always applied. A seam is introduced into the system Hamiltonian,
which continuously weakens existing bonds and simulaneously establishes new
bonds until the lattice is cut in two halfs of size $M^{d-1} \times L/2$.
Histograms taken in the seam energy give access to the change of the free
energy as a function of the seam strength \cite{KL96}, which finally yields
the total change of the free energy when the lattice is cut in two. For
periodic boundary conditions this method yields the Casimir amplitude
$\Delta_{per}$ directly. For other boundary conditions the knowledge of
$\Delta_{per}$ is required as an input information \cite{KL96}.
The method works very well for $d = 2$ for critical Potts models with
$q = 2$, 3, and 4 and has subsequently been applied to the Ising
model for $d = 3$ with periodic boundary conditions \cite{KL96} and with
surface fields \cite{MK97}. A summary of the currently available estimates
for the Casimir amplitudes from various sources is displayed in Table
\ref{Dtable} which includes older Migdal-Kadanoff estimates taken from
Ref.\cite{INW86}. Apart from the well known numerical uncertainties
regarding the extrapolation of the $\varepsilon$-expansion and the
Migdal-Kadanoff renormalization scheme the agreement between the estimates
for each of the amplitudes is encouraging. Especially for $\Delta_{++}$
and $\Delta_{+-}$, where the $\varepsilon$-expansion and the Migdal-Kadanoff
scheme are particularly unreliable, the other estimates are fairly consistent.
There are still some prospects to improve the Migdal-Kadanoff estimates also
for these cases, but final results are not yet available \cite{DD99}. It
would also be desirable to obtain estimates for the Casimir amplitudes from
a field theoretic calculation for $d = 3$ directly, but attempts in this
direction have not yet been made. The Monte - Carlo estimates for
$\Delta_{++}$ and $\Delta_{+-}$ are obtained by extrapolating the individual
data to inifinite lattice size \cite{MK97}. For $\Delta_{+-}$ this works
rather well, but for $\Delta_{++}$ substantial systematic uncertainties
remain and additional data for larger systems are required to obtain a
reliable extrapolation (see Figs.4 and 5 in Ref.\cite{MK97}). At present
local functional methods as set up in Ref.\cite{BU98} seem to provide the
most reliable estimates for $\Delta_{++}$ and $\Delta_{+-}$, because the
dimensional dependence of these amplitudes appears to be captured rather
well by the local free energy functional. Finally, we note that the Casimir
amplitudes may also be accessible by exploring the order parameter
distribution at the critical point \cite{Bruce95}. So far this method has
only been used for fully finite cubic geometries, generalizations to other
geometries have not yet been explored.

\subsection{Off-lattice models and wetting scenario}
A great drawback of the Monte - Carlo method introduced in Ref.\cite{KL96}
is, that it cannot be generalized to temperatures $T \neq T_{c,b}$. The method
is based on the measurement of free energy differences, which correspond to
linear combinations of the scaling functions at different scaling arguments
for $T \neq T_{c,b}$. Data of extremely high accuracy would be required
to disentangle the individual contributions to the measured free energy
difference. An alternative approach is to mimic the complete wetting scenario
(see Ref.\cite{SD88}) near the critical end point of the demixing transition
in a binary liquid mixture in a computer simulation \cite{WK98}. The order
parameter in this case is the concentration of the mixture ($N = 1$, Ising
universality class) rather than the density difference between liquid and gas,
which is usually taken as the order parameter near the liquid-vapor critical
point. In fact, temperature and pressure are adjusted such that the mixture
is in its vapor phase very close to liquid-vapor coexistence but {\em far away}
from the liquid-vapor critical point. The interplay between the interparticle
potential and the interaction between the particles and an external wall
(substrate) may lead to the formation of a macroscopic liquid wetting layer
of thickness $L$ on the substrate at some temperature $T_w$ below the
liquid-vapor critical point \cite{SD88}. The problem in the preparation of
such a complete wetting layer for a binary mixture is to find a system, i.e.,
parameter values for a simulation, such that the critical end point of the
demixing transitions is {\em inside} the complete wetting regime, where the
macroscopic wetting layer remains stable (see Ref.\cite{SD88} for more
background information on wetting transitions). The Casimir effect
associated with the critical demixing transition can then be studied in a
liquid layer of thickness $L$. The suggestion to probe the Casimir effect in
complete wetting layers near critical end points was first made by Nightingale
and Indekeu \cite{NI85} and was later worked out in more detail, as the first
estimates for the scaling functions of the Casimir force became available
\cite{KD92} (see also Sec.6 of Ref.\cite{MKbook}). The main objective of such
a simulation, however, is to obtain more insight into the Casimir effect in a
more realistic off-lattice model with Lennard - Jones interactions, which
would be the typical situation in an experiment \cite{WK98}. Simulations have
been performed for a symmetric binary liquid characterized by the Lennard -
Jones interparticle potential
\begin{equation}
\label{uij}
u_{ij}(r) = 4 \epsilon_{ij}\left[ \left({\sigma_{ij} \over r}\right)^{12}
- \left({\sigma_{ij} \over r}\right)^6 \right]
\end{equation}
for two paricle species $i,j = 1,2$, where the choices $\sigma_{ij} = \sigma$
for the hard core parameters and $\epsilon_{11} = \epsilon_{22} = \epsilon$,
$\epsilon_{12} = 0.7 \epsilon$ for the well depth parameters have been made.
Note that with these choices the system is invariant under the species
exchange $1 \leftrightarrow 2$ and therefore the chemical potentials of both
species have been set equal $\mu_1 = \mu_2 = \mu$ from the outset. The
simulations are performed in a box of size $P^2 \times D$, where periodic
boundary conditions are applied in the $x$ and $y$ directions and two hard
walls are specified in the $z$ direction, one at $z = 0$ and one at $z = D$.
The wall at $z = 0$ is characterized by the attractive potential
\begin{equation}
\label{Vz}
V(z) = \epsilon_w \left[ {2 \over 15} \left({\sigma \over z}\right)^9
- \left({\sigma \over z}\right)^3 \right]
\end{equation}
which acts {\em equally} on the particle species \cite{WK98}.
The interparticle interactions are truncated at $R_c = 2.5 \sigma$, whereas
no range cutoff is employed for $V(z)$. The phase diagram of the system is
spanned by the parameters $\mu / k_B T$, $\epsilon / k_B T$, and $\epsilon_w /
k_B T$. The system sizes used are $P = 12.5\sigma$, $15\sigma$, and
$17.5\sigma$, where $D = 40\sigma$ in all cases \cite{WK98}. The parameters
are chosen such that $i)$ a complete wetting layer forms on the substrate and
$ii)$ the endpoint of the critical demixing transitions is inside the
complete wetting regime (see above). In order to obtain a sufficiently thick
wetting layer $(L \geq 10\sigma)$ the undersaturation of the vapor $\delta
\mu / \mu$ must be tuned to about $10^{-3}$. The data aquisition is strongly
hamperd by large fluctuations of the liquid vapor interface position
(capillary waves) which also lead to a substantial slowing down of the
algorithm. Furthermore, the data are strongly affected by lateral finite-size
effects, because the capillary wave fluctuations also limit the lateral
system sizes attainable within reasonable computational effort \cite{WK98}.

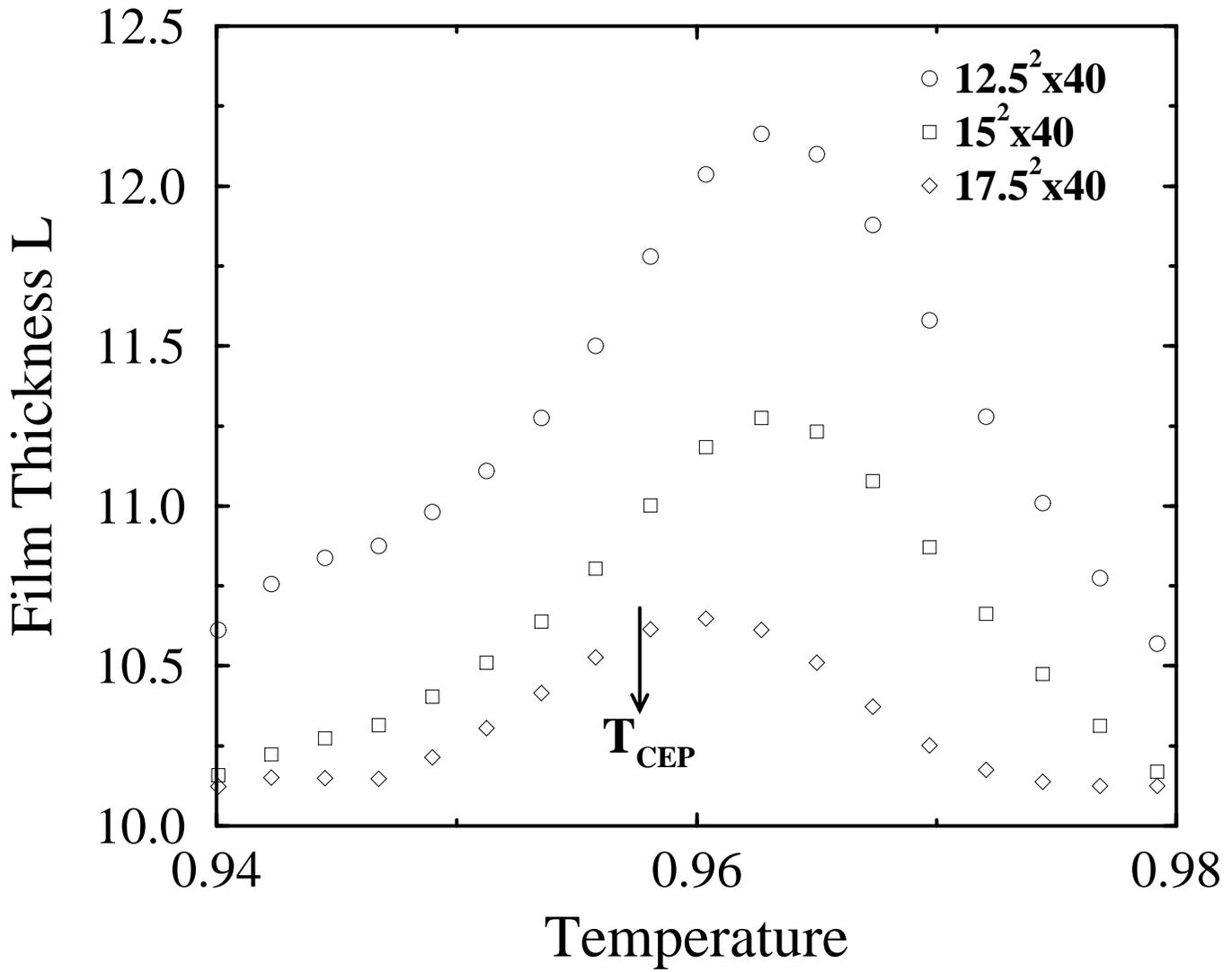
\begin{figure}[t]
\caption{Thickness of the wetting layer as a function of temperature in
reduced Lennard - Jones units along a path parallel to the liquid vapor
coexistence line. Data are shown for each of the three system sizes studied.
The results were obtained from multihistogram extrapolation of simulation
data accumulated at three points on this path, corresponding to temperatures
$T = 0.946$, 0.958, 0.97 (see Ref.\protect\cite{WK98}).
\label{LTplot}}
\end{figure}

The equilibrium thickness $L$ of the wetting layer (see also Sec.4) minimizes
the effective interface potential $\omega(l)$ \cite{SD88,WK98} as a function
of the test layer thickness $l$, which is a variational parameter in the
spirit of mean-field theory. The effective interface potential is given by
\begin{equation}
\label{oml}
\omega(l) = l (\rho_l - \rho_v) \delta \mu + \sigma_{wl} + \sigma_{lv} +
\delta \omega(l),
\end{equation}
where $\rho_l$ and $\rho_v$ denote the liquid and vapor densities,
respectively. The interfacial tensions $\sigma_{wl}$ between the wall
$(z = 0)$ and the liquid and $\sigma_{lv}$ between the liquid and the vapor
do not depend on $l$. The last term $\delta \omega(l)$ contains the
contributions of all interactions across the wetting layer and it depends
on the boundary conditions. By construction there are no surface fields
acting on the model liquid which break the $1 \leftrightarrow 2$ symmetry
between the particle species, i.e., ${\bf H}_1 = 0$ (see Eq.(\ref{Hs}) for
$N = 1$). However, the wall potental given by Eq.(\ref{Vz}) acts like a
{\em negative} surface enhancement $c$ (see Eq.(\ref{Hs})), which supports
demixing near the surface, whereas the liquid-vapor interface acts as a
free surface $(c > 0)$ due to the internal $1 \leftrightarrow 2$ symmetry
of the model liquid. The surface universality classes should thus be
characterized by the combination $(ab) = (O+)$. {\em At} $T = T_{cep}$
$\delta \omega(l)$ is therefore used in the form
\cite{WK98}
\begin{equation}
\label{doml}
\delta \omega(l) = {W \over l^2} + k_B T_{cep} \left({\Delta_{O+} \over l^2}
+ {2 l \Delta_{per} \over P^3} \right),
\end{equation}
where the Hamaker constant $W \simeq 2.5 k_B T_{cep}$ governs the van der
Waals contribution to the interaction potential and the Casimir amplitude
$\Delta_{O+} \simeq 0.2$ (see Table \ref{Dtable}) governs the Casimir
interaction for a symmetric liquid mixture \cite{WK98}. Note that positive
Casimir amplitudes (repulsion) lead to a critical {\em thickening} of the
wetting layer, whereas negative Casimir amplitudes (attraction) lead to a
critical {\em thinning} of the wetting layer. The last term in Eq.(\ref{doml})
is governed by the Casimir amplitude $\Delta_{per} \simeq -0.15$. It provides
an order-of-magnitude account of the aforementioned lateral finite-size
effects which can be treated as a shift of the undersaturation $\delta \mu$.
In the limit $P \gg l$ (film geometry), the critical thickening of the wetting
layer is given by \cite{WK98}
\begin{equation}
\label{LcL0}
L_c / L_0 = \left(1 + k_B T_{cep} \Delta_{O+} / W\right)^{1/3},
\end{equation}
where $L_c = L(T_{cep})$ and $L_0$ is the equilibrium layer thickness outside
the critical regime. The measured film thickness $L(T)$ versus temperature
is shown in Fig.\ref{LTplot}. The critical thickening of the wetting layer
$L_c/L_0 - 1$ for the largest system (lowest curve in Fig.\ref{LTplot}) is
of the order of 3\% which is in rough agreement with Eq.(\ref{LcL0}). The
apparent reduction of the critical thickening with incrasing lateral system
size can be explained semiquantitatively by the lateral finite-size
correction included in Eq.(\ref{doml}). Further studies of off-lattice models
like this are certainly desirable, however, algorithmic improvements for the
treatment of capillary waves will be indispensable for future progress.

\subsection{Lattice stress tensor}
The computer simulation of the complete wetting scenario is quite successful,
but by its design it only gives an indirect account of the scaling functions
$K$ of the Casimir force. As already described in Sec. 2 the most direct access
to the Casimir force is given by the thermal average of the {\em stress tensor}
and it would therefore be most convenient to have a lattice expression for
the stress tensor available for spin models. Such expressions can indeed be
obtained and successfully used in Monte - Carlo simulations for lattice models
for $d = 2$ \cite{BK98}. The basic idea behind the construction of the stress
tensor is the same is in continuum theory: one calculates the response of the
free energy to a nonconformal mapping of the system, e.g., an {\em anisotropic}
rescaling of the coupling constants. For example, one may choose $J_x = J_x
(\lambda)$, $J_y = J_y(\lambda)$ with $J_x(0) = J_y(0) = J$ and $J_x(\lambda)
= J_y(-\lambda)$ on a square lattice, such that the critical temperature does
not depend on $\lambda$. For the Ising model for $d = 2$ this procedure yields,
e.g., the $xx$ component of the lattice stress tensor at lattice site $(i,j)$
in the form \cite{BK98}
\begin{equation}
\label{txx}
t_{xx}(i,j) = -J_x'(0)\left( S_{i,j} S_{i+1,j} - S_{i,j} S_{i,j+1} \right),
\end{equation}
where $J_x'(0)$ is the derivative of $J_x(\lambda)$ at the isotropic point
$\lambda = 0$. The thermal averages $\langle t_{xx} \rangle$ of Eq.(\ref{txx})
and $\langle T_{xx} \rangle$ of the stress tensor in conformal field theory
are related by $\langle t_{xx} \rangle = \alpha \langle T_{xx} \rangle$ up to
corrections to scaling, where $\alpha$ is exactly known for the 2d Ising
model. For periodic boundary conditions in a strip geometry Eq.(\ref{txx})
can be used to measure the Casimir amplitude $\Delta_{per}$ (i.e., the
conformal anomaly number $c$) and two or more scaling dimensions for the
Ising model and other models if Eq.(\ref{txx}) is generalized accordingly
\cite{BK98}. For lattice models for $d = 2$ conformal field theory provides
sufficient background information so that the desired quantities can be
extracted from elaborate fit procedures \cite{BK98}. Although Eq.(\ref{txx})
can be readily generalized to $d = 3$, additional information from conformal
field theory, which is vital for the data interpretation
for $d = 2$, is no longer available. Furthermore, $\langle t_{xx} \rangle$
still contains surface contributions for nonperiodic boundary
conditions, because the surface tensions will depend on the anisotropy
parameter $\lambda$ even if the critical point does not. However, for
periodic boundary conditions $\langle t_{xx} \rangle$ is at least proportional
to the Casimir force and some preliminary studies for the XY model in an
$M^2 \times L$ geometry for $d = 3$ dimensions look promising \cite{MK99},
although high statistics is needed already for small systems.

\section{Experiments}
Experimental verifications of the Casimir effect in critical liquids are
exceedingly difficult, because data of high accuracy are required and
both samples and apparatus must be prepared with great care. At present,
two lines of approach are considered, namely the wetting scenario scetched
already in Sec.3 and direct force measurements by atomic force microscopes
(AFM).

\subsection{Wetting experiments}
For a wetting experiment in the vicinity of a critical point a fluid is
required which posesses a critical end point on the liquid vapor coexistence
line. One option for this setup is provided by $^4$He near its lower
$\lambda$ point \cite{GC99}. For this system the interaction part of the
effective interface potential (see Eqs.(\ref{oml}) and (\ref{doml})) must be
modified according to the universality class of the $\lambda$-transition in
$^4$He (XY, $N = 2$). The boundary conditions at the two interfaces of the
wetting layer seem to be very well approximated by Dirichlet boundary
conditions (O surface universality class). This leads to \cite{KD92,GC99}
\begin{equation}
\label{domlHe}
\delta \omega(l) = {W \over l^2} \left(1 + {l \over L_x}\right)^{-1}
+ {k_B T_\lambda \over l^2} \theta_{OO} \left(t\ l^{1/\nu},
\delta \mu\ l^{\beta \delta/\nu} \right),
\end{equation}
where $L_x \simeq 193 \AA$ denotes the crossover length to retardation
\cite{CC88}, and $\theta_{OO}$ is the scaling function of the Casimir
potential for the ordinary surface universality class (see Eq.(\ref{df})).
Note that $W$ and $L_x$ depend on the dielectric properties of the adsorbate
and the substrate. From Eqs.(\ref{oml}) and (\ref{domlHe}) one expects a
critical {\em thinning} of the wetting layer thickness, because $\theta_{OO}
< 0$. Note that the second scaling argument of $\theta_{OO}$ captures
off-coexsistence effects due to the undersaturation $\delta \mu$ of the
$^4$He vapor. At the $\lambda$ point $(t = 0, \delta \mu = 0)$ one has
$\theta_{OO}(0,0) = \Delta_{OO} \simeq -0.024$ which results in a critical
thinning of $\sim 0.3\%$ for standard substrates like, e.g., copper
\cite{KD92,MKbook}. In the experimental setup a stack of five copper
capacitors is placed inside a cell which contains liquid $^4$He at the bottom.
The surfaces of the capacitor provide the substrate potential (see
Eq.(\ref{Vz})) and their elevation $h$ in the gravitational field controls the
undersaturation $\delta \mu \sim \rho_v g h$ of the $^4$He vapor. The layer
thickness is obtained from high precision measurements of the capacitance of
each of the capacitors as a function of temperature. The wetting behavior of
$^4$He is extremely sensitive to the surface morphology of the copper plates.
In particular, microscopic scratches and dust particles lead to localized
condensation of $^4$He on the surface which results in an overestimation of
the thickness. Moreover, surface roughness leads to an enhanced surface area
which also increases the amount of liquid $^4$He on the substrate. Even with
the most advanced polishing techniques these effects cannot be avoided
completely and therefore also the experimental verification of the DLP theory
of dispersion forces \cite{Dispersion} remains a challenge. Nevertheless,
the experimental data of the film thinning show a pronounced minimum well
below $T_\lambda$ which is given by the specific value $x_m = -9.2 \pm 0.2$
of the first scaling argument $x = t L^{1/\nu}$ in Eq.(\ref{domlHe}). This
value coincides with the minimum of the scaling function $K_{OO}(x,y)$ of the
Casimir force with respect to $x$. The experimental estimate of $\vartheta(x)
\equiv K_{OO}(x,y=0)$ extracted from the data is displayed in Fig.\ref{Kxplot},
which does not show the expected data collapse for the scaling function
$K_{OO}$. On the contrary, a systematic trend in the data as function of the
elevations $h$ of the capacitors is visible as shown in the inset of
Fig.\ref{Kxplot}. One possible explanation may be given by off-coexistence
effects, which would require the second scaling variable $y = \delta \mu\
L^{\beta \delta/\nu}$ in Eq.(\ref{domlHe}) for data collapse.
If a linear dependence of $K_{OO}$ on $y$ is assumed, the deviations from
data collapse are indeed drastically reduced \cite{GC99}. Another option is
provided by the introduction of a roughness correction factor as suggested in
Ref.\cite{Li92}, which leads to a similar improvement \cite{GC99}. However,
it is evident from Fig.\ref{Kxplot}, that there are no appreciable deviations
from data collapse for $x \geq 0$, where a quantitative prediction for
$K_{OO}(x,y=0)$ exists \cite{KD92,MKbook}. The comparison is displayed in
Fig.\ref{Kxscal} which shows reasonable agreement between the data and the
prediction. Finally, we note that a finite thinning effect remains visible
for temperatures further below $T_\lambda$, as one would expect from the
presence of Goldstone modes \cite{Li92,Danchev98}. The overall shape
of $K_{OO}$ is manifestly nonmonotonic in contrast to the recently stated
monotonicity hypothesis for $O(N>1)$ symmetric spin models \cite{Danchev98}.

As second option for a wetting experiment in the vicinity of a critical end
point is provided by binary liquid mixtures near the critical end point
$T_{cep}$ (see Sec.3) of the line of second order demixing transitions
\cite{ML99} (see Refs.\cite{SD88,KD92} for complete phase diagrams). The
physical situation is very much like $^4$He near the lower $\lambda$-point,
except that both the bulk and surface universality classes are different
here. The second order demixing transition is characterized by a scalar
order parameter (concentration), so the system is in the Ising $(N = 1)$
universality class. The substrate material as well as the liquid - vapor
interface, which provide the boundaries of the system, usually show some
preferential affinity for one of the two components of the mixture, i.e.,
the concentration (order parameter) departs from its critical bulk value
in the vicinity of the surfaces. This phenomenon is known as critical
adsorption (see, e.g., Ref.\cite{SurfRev}) and it is captured by the
{\em extraordinary} surface universality class. In the experiment \cite{ML99}
a molecularly smooth (100) Si wafer (n-type, phosphorous doping) covered
with a SiO$_2$ layer of $\sim 2.0 nm$ thickness is used, which is suspended
vertically inside a pyrex cell. The elevation $h$ of the substrate above the
bulk liquid, at which the wetting layer thickness is measured, controls the
undersaturation of the vapor. The reduction of temperature gradients for
wetting agents other than superfluid $^4$He is a quite demanding task and
it substantially complicates the preparation of the cell and the sample
\cite{ML99}. In this experiment two organic mixtures have been used, namely
methanol + hexane (MH) and 2-methoxy-ethanol + methylcyclohexane (MM).
In MH the methanol component is adsorbed at the Si wafer, whereas hexane
is adsorbed at the liquid - vapor interface of the wetting layer. In MM the
situation is similar: the 2-methoxy-ethanol is adsorbed at the Si wafer,
whereas the methylcyclohexane is adsorbed at the liquid - vapor interface.
The wetting layers of both mixtures are therefore characterized by the
scaling functions $\theta_{+-}$ of the Casimir potential. The interaction
contribution $\delta \omega(l)$ to the effective interface potential
$\omega(l)$ in this case is assumed to be of the form \cite{ML99}
\begin{equation}
\label{domlBL}
\delta \omega(l) = {W \over l^2} - A e^{-l / d}
+ {k_B T_{cep} \over l^2} \theta_{+-} \left(t\ l^{1/\nu}, 0\right),
\end{equation}
where retardation and off - coexistence effects are neglected. The exponential
contribution to Eq.(\ref{domlBL}) is due to the presence of the hard wall,
which structures the adsorbed fluid over a molecular distance $d$. The
critical temperature $T_{cep}$ is about $300 K$ \cite{ML99}. At the
moment only mean field results \cite{MK97} and exact results for $d = 2$
\cite{ES94} exist for the scaling function $K_{+-}$ of the Casimir force.
In order to obtain reasonable estimates also for $d = 3$ at least the one -
loop corrections are required which only exist for $\Delta_{+-} = 2
K_{+-}(0,0)$ at the moment (see Ref.\cite{MK97} and Table \ref{Dtable}).
From Eq.(\ref{LcL0}) and typical values for $T_{cep}$ and the Hamaker constant
$W$ one expects a critical thickening $L_c/L_0 \geq 2$ of the wetting layer,
when the estimate $\Delta_{+-} \simeq 2.4$ (see Table \ref{Dtable}) is used.
As $T_{cep}$ is approached from above a critical thickening of the wetting
layer consistent with this expectation has been found in the experiment and
the data for $K_{+-}(x,0)$ are indeed consistent with scaling \cite{ML99}.
As function of the scaling variable $y \equiv L/\xi_+$ the scaling function
$\vartheta_{+-}(y) \equiv K_{+-}(x = (\xi_0^+ y)^{1/\nu},0)$ is shown in
Fig.\ref{Kpmscal}. A comparison between $\vartheta_{+-}(y)$ {\em at} and
away from the critical composition is shown in the inset for MH. The shape
of $\vartheta_{+-}(y)$ shown in Fig.\ref{Kpmscal} resembles that of the
mean field estimate $K_{+-}(x,0)$ for $x > 0$ shown in Fig.\ref{Kx0}.
However, the Hamaker constant $W$ and therefore also the Casimir amplitude
$\Delta_{+-}$, which have been extracted from the data, are much smaller
than anticipated. The reason for this discrepancy has not yet been fully
understood. One possible explanation could be that the parameters of the
system are not in the complete wetting regime, i.e., only {\em partial}
rather than complete wetting \cite{SD88} is achieved. Further studies are
currently under way.

\subsection{AFM measurements}
As already mentioned in Sec.3 the theoretical investigation of critical
fluids in curved geometries is, inter alia, motivated by the prospects
of measuring the Casimir force directly with an AFM. At the moment only
exploratory results are available \cite{IM96} which have been obtained for
the SPW geometry in liquid crystals (see Refs. \cite{APP91,Ziherl98}). In
this study a temperature controlled AFM has been used to measure the force
between a sphere mounted on the cantilever tip of the AFM and a planar wall
immersed in an 8CB liquid crystal near the isotropic - nematic phase
transition. Above the transition in the isotropic phase an attractive
force of the order of $10^{-10} N$ at a distance $D = 1nm \pm 0.1nm$ between
the surface of the sphere $(R = 5\mu m)$ and the wall is detected only when
the two surfaces are {\em moving apart}. This phenomenon is similar to the
capillary force in AFM microscopy and it is interpreted as the adsorption of
a nematically ordered layer of the liquid crystal on the surface of the
sensing probe \cite{IM96}. Slightly above the transition to the
nematically ordered phase an additional attractive force of the order of
$10^{-11}N$ is detected when the surfaces are {\em approaching} one another.
This additional force is conjectured to be the Casimir force mediated by
the onset of Goldstone modes of the nematic director field in the ordered
phase \cite{APP91,IM96}, where the boundary conditions are supplied by the
type of anchoring of the nematic director on the surfaces \cite{APP91}.
Further quantitative studies of Casimir forces in critical and correlated
liquids with this apparatus are certainly desirable. Finally, we note that
the radiation pressure on a dielectric sphere in the evanescent field of
totally reflected light has recently been measured using such AFM techniques
\cite{VMC98}.

\section{Prospects for further investigations}
The theoretical knowledge about Casimir forces in critical and correlated
fluids which has been accumulated during the last 10 years has become
so detailed, that the stage is set for experimental tests of various kinds.
Wetting experiments near critical end points have already proved to be a
powerful tool to accomplish this goal for quite a variety of fluids.
Further studies in this direction are certainly highly desirable and the
prospects for them are very good despite substantial experimental challenges
one has to face. From the existing theoretical work on curved geometries it
has also become clear, that the Casimir forces in critical and correlated
fluids are within reach of current AFM designs. Preparation of the samples
and temperature stabilization of the sample and the instrument again pose
major challenges for AFM force measurements, however, the prospects of
probing the Casimir effect quantitatively are also very good.

Conversely, the experimental approaches to the Casimir effect also pose
new theoretical challenges. The problem of substrate roughness in wetting
experiments has already been mentioned above and for the case of quenched
roughness theoretical results already exist \cite{Li92}. However, one of
the boundaries of a wetting layer is a free liquid - vapor interface,
which may undergo large scale fluctuations due to capillary waves. What kind
of corrections capillary waves as additional degrees of freedom impose on
the critical Casimir potential is an open question. Quantitative estimates
of these corrections are not only important for experiments, they would also
aid the data interpretation of computer simulations for critical wetting
layers. To what extent off-coexistence effects influence experimental and
numerical wetting layer data is also a largely open problem. In this respect
the recently explored numerical access to the stress tensor of lattice
models may prove particularly useful \cite{BK98,MK99}. Finally, it
should be mentioned that improvements of existing theoretical or numerical
estimates for the scaling function $K_{ab}$ of the Casimir force in particular
for the extraordinary surface universality class in various geometries are
still needed in order to extract the Casimir effect from experimental data
as reliably as possible.

Although the history of the Casimir effect goes back more than half a century
it has remained an active field of research. This article can therefore only
provide a snapshot of current knowledge in this area rather than a completed
picture. If this presentation could finally help to trigger or direct new
research work in this field, then its main purpose would be fulfilled.

\section*{Acknowledgments}
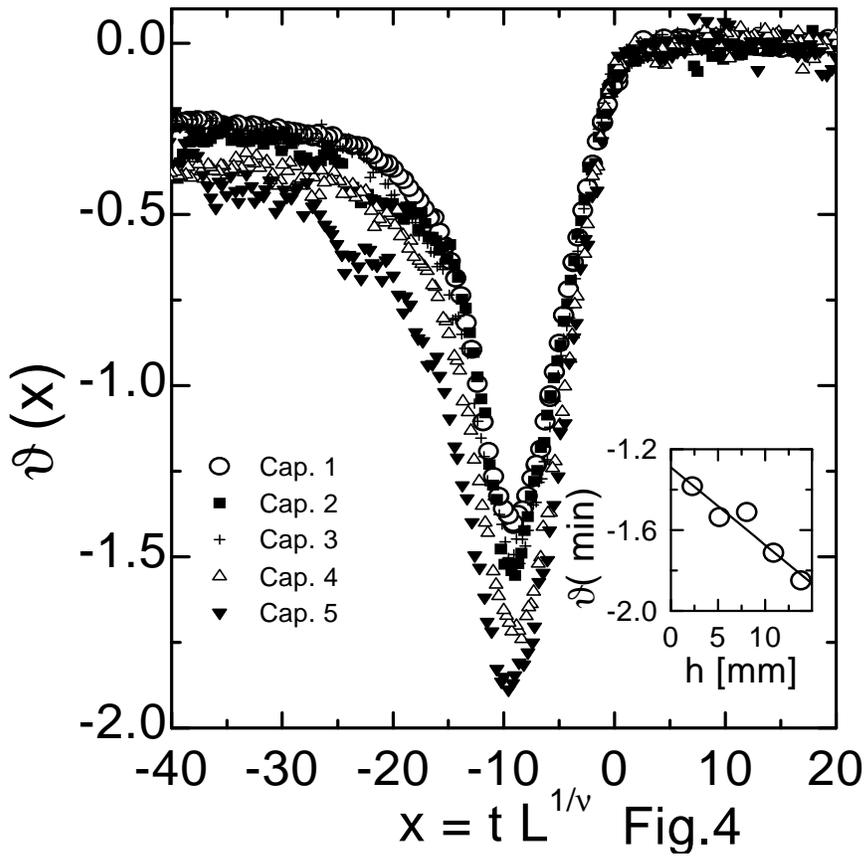
\begin{figure}[t]
\caption{Scaling function $\vartheta(x) \equiv K_{OO}(x,0)$ as a function
of $x$. The magnitude of the minimum increases systematically with the height
$h$ of the capacitor. The measured layer thickness $L$ is roughly between
300\AA\ and 600\AA\ depending on the capacitor index 1 -- 5. The inset shows
the value of $\vartheta(x)$ at the minimum vs. height. The uncertainty in the
vertical scale is 2 - 10\% (taken from Ref.\protect\cite{GC99}).
\label{Kxplot}}
\end{figure}

The author gratefully acknowledges stimulating discussions with M. H. W.
Chan, S. Dietrich, R. Garcia, M. Kardar, B. M. Law, A. Mukhopadhyay, and
F. Schlesener. Financial support for this work has been provided through
the Heisenberg program of the Deutsche Forschungsgemeinschaft, which is
also gratefully acknowledged. The author expresses his special thanks to
M. H. W. Chan and R. Garcia for providing copies of Figs.\ref{Kxplot} and
\ref{Kxscal} and to B. M. Law and A. Mukhopadhyay for providing a copy
of Fig.\ref{Kpmscal} prior to publication.

\begin{figure}[t]
\caption{Blow-up of the region $x \geq 0$ in Fig.\protect\ref{Kxplot}.
Every other data point is shown. The solid line shows the prediction
from Fig.9 in Ref.\protect\cite{KD92} (taken from Ref.\protect\cite{GC99}).
\label{Kxscal}}
\end{figure}

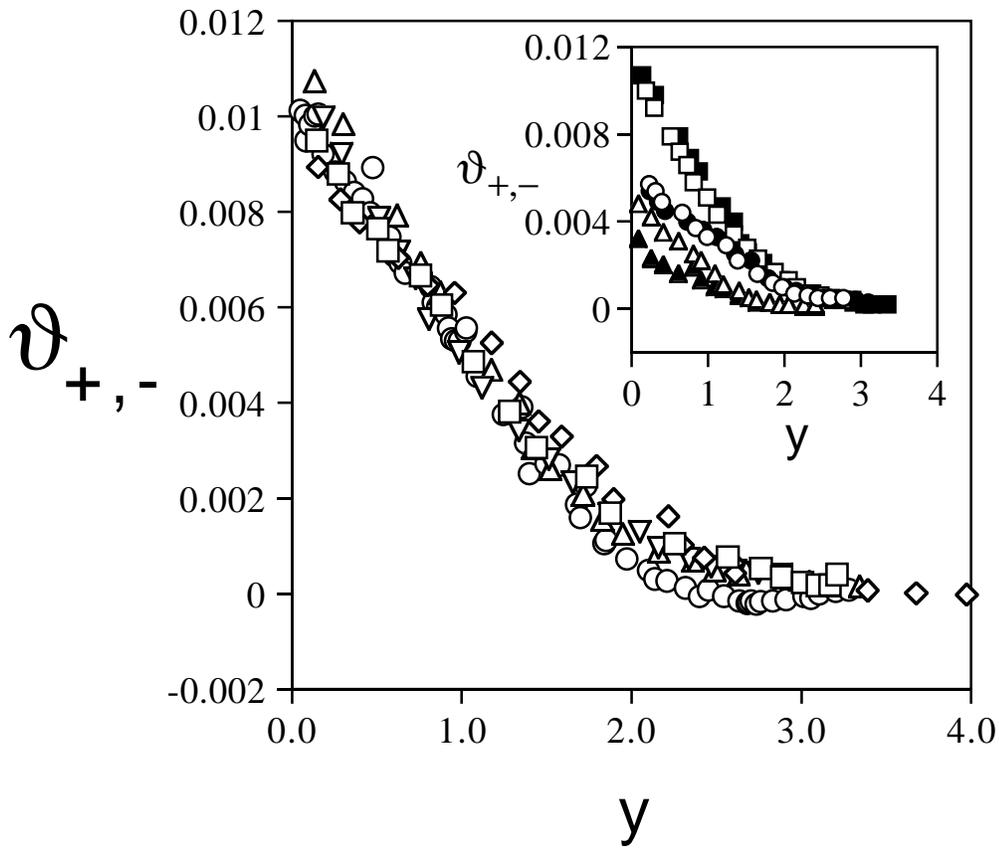
\begin{figure}[t]
\caption{Universal scaling function $\vartheta_{+,-}(y)$, $y = L/\xi_+$ for
the critical Casimir force. The symbols represent data at fixed elevations
$h = 1.5mm$ (diamonds), $h = 3.3mm$ (squares) for MM and $h = 3.4mm$
(triangles), $h = 6.3mm$ (inverted triangles) for MH. In the inset the
experimental $\vartheta_{+,-}(y)$ for the system MH with critical composition
(squares), 5\% excess hexane (circles) and 10\% excess hexane (triangles) at
two different heights (3.5 mm (open symbols) and 6.0 mm (solid symbols)) on a
silicon wafer is shown.
\label{Kpmscal}}
\end{figure}

\begin{table}[t]
\caption{
Casimir amplitudes for the Ising universality class for $d = 3$. The
values labelled $\varepsilon = 1$ are obtained by extrapolating
the $\varepsilon$-expansion for $N = 1$ to $\varepsilon = 1$
\protect\cite{MK97}. The values labelled $d = 3$ are obtained from
Pade type approximants for $d = 3$ $(\varepsilon=1)$ \protect\cite{MK97}.
The Monte-Carlo estimates obtained from the algorithm presented in
Ref.\protect\cite{KL96} are labelled by 'MC'. Statistical errors (one
standard deviation) are indicated by the figures in parenthesis. The last
two lines show estimates taken from Refs.\protect\cite{INW86} and
\protect\cite{BU98}.
\label{Dtable}}
\end{table}

\begin{tabular}{lllllll}
\hline \hline
& $\ \ \Delta_{per}$ & $\ \ \Delta_{O,O}$ & $\ \ \Delta_{+,+}$ &
$\ \Delta_{+,-}$ & $\ \Delta_{SB,+}$ & $\ \Delta_{O,+}$ \\
\hline
$\varepsilon = 1$ & $-0.1116$ & $-0.0139$ & $-0.173$ & $1.58$ &
$-0.093\ \ \ $ & $0.165$ \\
$d = 3$ & $-0.1315$ & $-0.0164$ & $-0.326$ & $2.39$ & & $0.208$ \\
MC & $-0.1526(10)\ $ & $-0.0114(20)\ $ & $-0.345(16)\ $ &
$2.450(32)\ $ & & $0.1873(70)\ $ \\
Ref.\cite{INW86} & & $-0.015$ & $\ \ 0$ & $0.279$ & $\ \ \ 0.017$ &
$0.051$ \\
Ref.\cite{BU98} & & & $-0.428$ & $3.1$ & & \\
\hline \hline
\end{tabular}

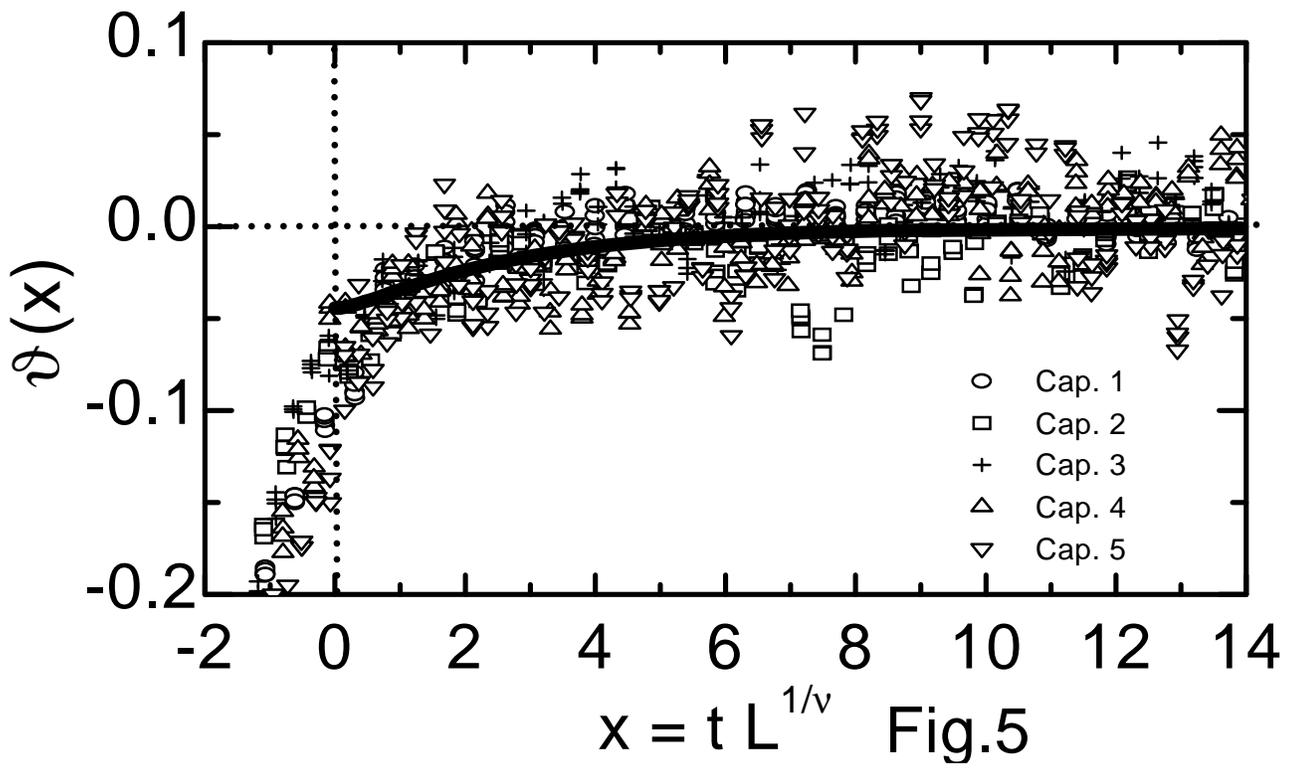
\end{document}